\documentclass[runningheads]{llncs}
\usepackage{graphicx}
\usepackage{xcolor}
\usepackage{multirow}

\newcommand{\ie}{\emph{i.e.}, }
\newcommand{\eg}{\emph{e.g.}, }

\begin{document}
\title{Opportunistic Screening of Osteoporosis Using Plain Film Chest X-ray}
%


\author {
 \thanks{This work was done when Fakai Wang interned at PAII Inc.}
 Fakai Wang\textsuperscript{\rm 1,\rm 2},
 Kang Zheng\textsuperscript{\rm 1},
 Yirui Wang\textsuperscript{\rm 1},
 Xiaoyun Zhou\textsuperscript{\rm 1},
 Le Lu\textsuperscript{\rm 1},
 Jing Xiao\textsuperscript{\rm 4 },
 Min Wu\textsuperscript{\rm 2 },
 Chang-Fu Kuo\textsuperscript{\rm 3 },
 Shun Miao\textsuperscript{\rm 1 } \\
 }
\institute{
\textsuperscript{\rm 1} PAII Inc., Bethesda, Maryland, USA  \\ 
 \textsuperscript{\rm 2} University of Maryland College Park, USA \\
\textsuperscript{\rm 3} Chang Gung Memorial Hospital, Linkou, Taiwan, ROC \\
\textsuperscript{\rm 4} Ping An Technology, Shenzhen, China
}

\maketitle              
\begin{abstract}

%

%
Osteoporosis is a common chronic metabolic bone disease that is often under-diagnosed and under-treated due to the limited access to bone mineral density (BMD) examinations, \eg Dual-energy X-ray Absorptiometry (DXA). In this paper, we propose a method to predict BMD from Chest X-ray (CXR), one of the most common, accessible and low-cost medical image examinations. Our method first automatically detects Regions of Interest (ROIs) of local and global bone structures from the CXR. Then a multi-ROI model is developed to exploit both local and global information in the chest X-ray image for accurate BMD estimation. Our method is evaluated on 329 CXR cases with ground truth BMD measured by DXA. The model predicted BMD has a strong correlation with the gold standard DXA BMD (Pearson correlation coefficient \textit{0.840}). When applied for osteoporosis screening, it achieves a high classification performance (AUC \textit{0.936}). 
As the first effort in the field to use CXR scans to predict the spine BMD, the proposed algorithm holds a strong potential in enabling early osteoporosis screening through routine chest X-rays and contributing to the enhancement of public health.

\end{abstract}
\section{Introduction}

Osteoporosis is the most common chronic metabolic bone disease, characterized by low bone mineral density (BMD) and decreased bone strength to resist pressure or squeezing force. With an aging population and longer life span, osteoporosis is becoming a global epidemic, affecting more than 200 million people worldwide~\cite{sozen2017overview}. Osteoporosis increases the risk of fragility fractures, which is associated with reduced life quality, disability, fatality, and financial burden to the family and the society. While with an early diagnosis and treatment, osteoporosis can be prevented or managed, osteoporosis is often under-diagnosed and under-treated among the population at risk~\cite{lewiecki2019challenges}. More than half of insufficiency fractures occur in individuals who have never been screened for osteoporosis ~\cite{smith2019screening}. The under-diagnosis and under-treatment of osteoporosis are mainly due to 1) low osteoporosis awareness and 2) limited accessibility of Dual-energy X-ray Absorptiometry (DXA) examination, which is the currently recommended modality to measure BMD. 

Opportunistic screening of osteoporosis is an emerging research field in recent years~\cite{cheng2020opportunistic,dagan2020automated,jang2019opportunistic,pickhardt2020automated}. It aims at using medical images done for other indications to screen for osteoporosis, which offers an opportunity to increase the screening rate at no additional cost and time. Previous attempts mainly focused on using the CT attenuation (i.e., Housefield Unit) of the vertebrae to correlate with BMD and/or fracture risk. Plain films have much greater accessibility than CT scans. Its excellent spatial resolution permits the delineation of fine bony microstructure that may contain information that correlates well with the BMD.  We hypothesize that specific regions of interest (ROI) in the standard chest x-rays (CXR), the most common medical imaging prescribed clinically, may help the opportunistic screening for osteoporosis. 

In this work, we introduce a method to estimate the BMD from CXR and screen osteoporosis. Our method first locates anatomical landmarks of the patient's bone structures and extracts multiple ROIs that may provide imaging biomarkers for osteoporosis. We proposed a novel network architecture that jointly processes the ROIs to accurately estimate the BMDs. We evaluate our method on 1651 CXRs from 1638 patients with ground truth BMDs measured using DXA. The results demonstrate that the BMDs estimated from CXR achieve a high correlation with the DXA BMD (\ie r-value 0.840) and a high osteoporosis diagnosis AUC of 0.936. 

In summary, our contributions are three-fold: 1) we develop the first method to estimate BMD from CXR, leading to a practical solution for opportunistic screening of osteoporosis; 2) we propose a clinical knowledge-driven method to combine global and local information in CXR to accurately estimate lumbar spine BMD. 3) we demonstrate that our method has significant performance gain over the baseline methods and achieves clinically useful osteoporosis screening performance.

\section{Methodology}
\subsection{Problem Overview}
Our task is to predict the BMDs of the patient's lumbar spine from a frontal chest X-ray. Formulated as a regression problem, the input is a CXR, and the output is the predicted BMD value. We choose to predict the BMDs of the lumbar spine because it is one of the standard region for BMD examination and is often affected by insufficiency fracture. Since the lumbar spine is not visible in CXR, a major challenge in this task is to identify the regions and visual patterns in the CXR that are correlated with the lumbar spine BMD. To overcome this challenge, our method comprises two steps: 1) extraction of Regions of Interest (ROIs) that are potentially correlated with the lumbar spine BMDs (described in Section \ref{ssec:landmarkandROI}); 2) training a multi-task regression neural network that jointly analyzes the ROIs to predict BMDs (described in Section \ref{ssec:fusion_model}).

\begin{figure*}[bt!]
	\centering
	\includegraphics[width=0.95\linewidth]{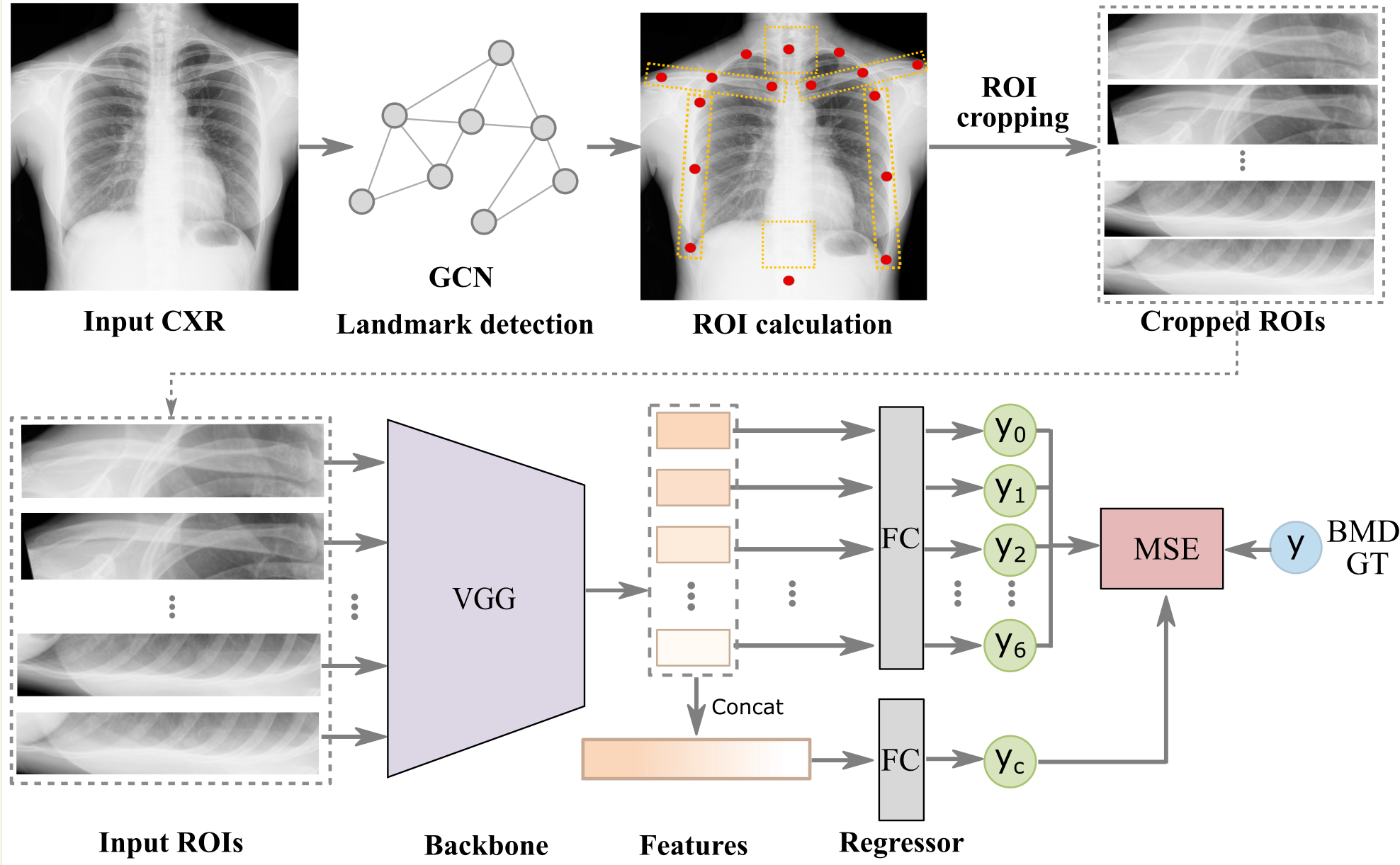}
	\caption{An overview of the training workflow, with landmark detection and ROI extraction in the upper part, and a multi-ROI model in the lower part. The landmark detector localizes 16 red dots on CXR images, to anchor local bones. Each dashed yellow box is normalized to the same orientation, height/width ratio. The cropped ROIs are used by all models. In the multi-ROI model, independent inputs include 6 local ROIs and the whole chest image. During training, Mean Square Error loss (MSE) is applied on 7 individual feature BMD predictions $y_0,..,y_6$, and on the concatenated feature BMD prediction $y_c$.}
	\label{fig:system}
\end{figure*}

\subsection{Automatic ROI Extraction}
\label{ssec:landmarkandROI}

Since it is unclear which local regions within the chest X-ray are the most informative for our task, we need to explore the correlations between the lumbar spine BMD and different regions in the CXR by extracting multiple ROIs. To exploit both local textures and global structures, the extracted ROIs should contain different scales. A clinical prior knowledge is that osteoporosis is a metabolic bone disease that affects all bones in the human body. Therefore, we hypothesize that bone regions in the CXR provide visual cues of the lumbar spine BMDs and extract ROIs from the following regions: left/right clavicle bones, cervical spine, left/right rib-cage area, T12 vertebra. Besides these 6 ROIs of local bone structures/patterns, we also include the whole CXR as another ROI to provide global structural information. To automatically extract the ROIs, we employ a graph-based landmark detection method, DAG~\cite{li2020structured}, and detect 16 landmarks which include 1) 3 points on the left/right clavicles, 2) 4 points along the left/right rib cages, 3) 1 point on the cervical spine and 1 point on the T12 vertebra. The ROIs are placed based on the detected landmarks. Examples of the CXR with 16 landmarks and extracted ROIs are shown in Figure~\ref{fig:system}. 

%

\subsection{BMD Estimation via Joint Analysis of the ROIs}
\label{ssec:fusion_model}

Given the 7 extracted ROIs, we proposed a multi-task learning network architecture, termed multi-ROI model, to jointly analyze the ROIs to estimate the lumbar BMDs. By jointly analyze the global and local ROIs, the network is capable of extracting visual patterns at different scales that are informative for predicting lumbar BMDs. In particular, we employ VGG-16 as the backbone network to produce a feature map encoding of the input ROI and apply Global Average Pooling (GAP) to aggregate the feature map to a single feature vector. The 7 ROIs go through the same VGG-16 network independently, generating 7 feature vectors. These feature vectors are then decoded both separately and jointly. First, they are individually processed by a shared fully connected (FC) layer to produce BMD estimations. Second, the 7 features are connected to obtain a joint feature vector, which is decoded by an FC layer to produce BMD estimation. During training, the mean squared error (MSE) is calculated on both the BMDs produced from the features of individual ROI and on the BMDs produced from the concatenated features. The average MSE is used as the training loss. During testing, the BMDs produced from the concatenated feature is taken as the output.

\vspace{-2mm}
\section{Experiments}
\label{sec:Experiments}

\vspace{-1mm}
\subsection{Experiment Setup}
\noindent{\bf Dataset.} We collected 1651 frontal view CXR scans with paired BMD scores (on four lumbar vertebrae L1 - L4) from \textit{anonymous hospital} after de-identification of the patient information. All experiments use the same data split, with 1087, 265 and 329 cases for training, validation and testing, respectively. The scans of the same patient only go to the same set.

\begin{figure*}[bt!]
	\centering
	\includegraphics[width=0.95\linewidth]{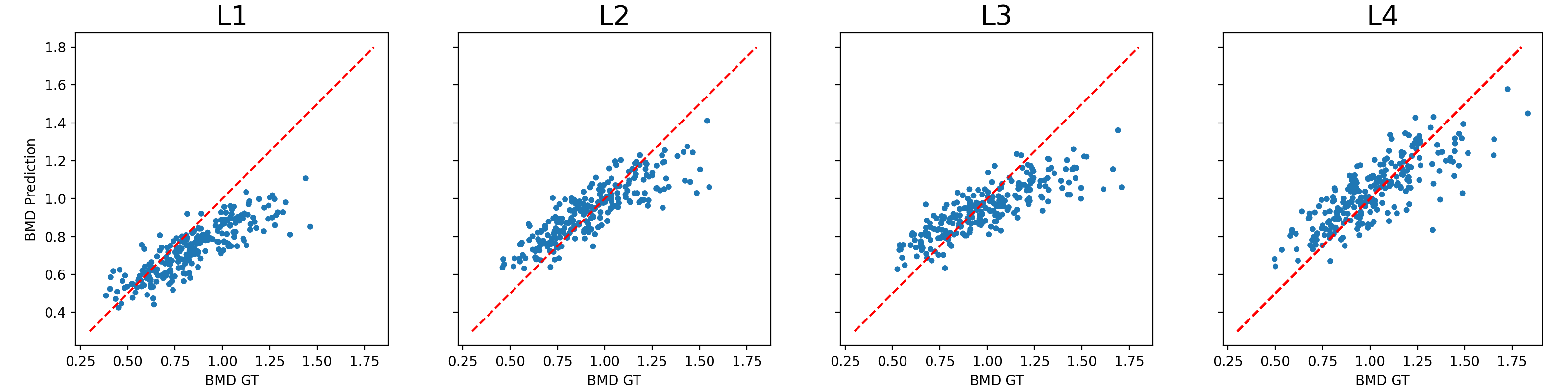}
	\caption{Multi-ROI model BMD prediction on different lumbar vertebra (L1-L4). X-axis is the ground truth BMD, and Y-axis represent the predictions.}
	\label{fig:fusionFClayerPredPoints}
\end{figure*}

\vspace{2mm}
\noindent{\bf Metrics.} We adopt two commonly used evaluation metrics: \textit{Pearson Correlation Coefficient} (R-value) and \textit{Area Under Curve}. R-value measures the linear relationship between the predicted BMD and the ground truth BMD. It only considers the correlation between two sequences regardless their absolute values. While Area Under Curve (AUC) measures accumulated true positive (osteoporosis) rate under different judging threshold for binary classification problem. We calculate the AUC score under the T-score range instead of the BMD values as T-score is a widely adopted measurement in the clinical practices. We convert the BMD scores to the T-score by checking the transforming table coming with the DXA machine. To evaluate osteoporosis diagnosis performance, we label the cases with ground truth T-scores below $-2.5$ as positive osteoporosis (clinical definition of osteoporosis).

\subsection{Implementation Details}
We trained our model on a workstation with Intel Xeon CPU E5-2650 v4 CPU @ 2.2 GHz, 132 GB RAM, and 4 NVIDIA TITAN V GPUs. Our models are implemented with PyTorch. X-ray inputs are resized to \textit{(256, 256) }, and repeated 3 times to form the RGB channels. During training, we apply augmentations such as scaling, rotation, translation, random flip. The SGD optimizer has a learning rate of $0.0001$, a weight decay of 4e-4. A mini-batch size of 64 is used to train the model for $100$ epochs.

\begin{table*}[]
	\caption{Comparison of different deep learning backbones. They all use the whole chest X-ray images as input, and use the baseline working flow. The only difference is convolutional backbone.}
	\centering

\begin{tabular}{|c|c|c|c|c|c|c|c|c|c|c|}
\hline
\multirow{2}{*}{Backbone} & \multicolumn{2}{c|}{L1}         & \multicolumn{2}{c|}{L2}         & \multicolumn{2}{c|}{L3}         & \multicolumn{2}{c|}{L4}         & \multicolumn{2}{c|}{Average}    \\ \cline{2-11} 
                          & R-value        & AUC            & R-value        & AUC            & R-value        & AUC            & R-value        & AUC            & R-value        & AUC            \\ \hline\hline
Vgg11                     & 0.843          & 0.919          & 0.838          & \textbf{0.916} & \textbf{0.832} & 0.917          & 0.772          & 0.881          & 0.821          & 0.908          \\ \hline
Resnet18                  & 0.759          & 0.875          & 0.799          & 0.863          & 0.774          & 0.852          & 0.717          & 0.840          & 0.762          & 0.858          \\ \hline
Resnet34                  & 0.787          & 0.903          & 0.787          & 0.866          & 0.768          & 0.887          & 0.723          & 0.861          & 0.766          & 0.879          \\ \hline
Resnet50                  & 0.799          & 0.894          & 0.823          & 0.888          & 0.795          & 0.880          & 0.741          & 0.858          & 0.789          & 0.880          \\ \hline
Vgg16                     & \textbf{0.865} & \textbf{0.923} & \textbf{0.863} & 0.913          & 0.824          & \textbf{0.918} & \textbf{0.804} & \textbf{0.901} & \textbf{0.839} & \textbf{0.914} \\ \hline
\end{tabular}

 \label{tab:comparisons}
\end{table*}

\subsection{Comparison of Backbone Models}
\label{ssec:baselinemodel}

We evaluate the performance of different backbone models using a baseline method. The baseline method takes the whole CXR as input. It employs a backbone network as encoder, followed by a GAP layer and a FC layer for regression. We evaluated multiple popular backbone networks and found that VGG16 delivers the best performance, as summarized in table~\ref{tab:comparisons}. Resnet provides complex feature computation and works better on classification tasks. The bone texture encoding task involves only low-level pattern recognition, where VGG models fit better. 

\begin{table*}[]
	\caption{Comparison of different modeling schemes. The fist 5 rows are baseline experiments on different ROIs (ROI models), the \textit{6th} is the multi-ROI scheme fed on all input modalities (multi-ROI model).}
	\centering
	\begin{tabular}{|c|c|c|c|c|c|c|c|c|c|c|}
\hline
\multirow{2}{*}{Modality} & \multicolumn{2}{c|}{L1}         & \multicolumn{2}{c|}{L2}         & \multicolumn{2}{c|}{L3}         & \multicolumn{2}{c|}{L4}         & \multicolumn{2}{c|}{Average}    \\ \cline{2-11} 
                          & R-value        & AUC            & R-value        & AUC            & R-value        & AUC            & R-value        & AUC            & R-value        & AUC            \\ \hline\hline
cervi                     & 0.823          & 0.915          & 0.817          & 0.930          & 0.802          & 0.937          & 0.738          & 0.841          & 0.795          & 0.906          \\ \hline
clavi                     & 0.793          & 0.893          & 0.778          & 0.917          & 0.745          & 0.867          & 0.709          & 0.857          & 0.756          & 0.883          \\ \hline
lumbar                    & 0.750          & 0.829          & 0.791          & 0.889          & 0.754          & 0.859          & 0.711          & 0.841          & 0.751          & 0.855          \\ \hline
ribcage                   & 0.748          & 0.868          & 0.768          & 0.876          & 0.730          & 0.850          & 0.703          & 0.830          & 0.737          & 0.856          \\ \hline
chest                     & 0.865          & 0.923          & \textbf{0.863} & 0.913          & 0.824          & 0.918          & \textbf{0.804} & 0.901          & 0.839          & 0.914          \\ \hline \hline
Prop.                    & \textbf{0.875} & \textbf{0.936} & 0.857          & \textbf{0.944} & \textbf{0.843} & \textbf{0.941} & 0.783          & \textbf{0.924} & \textbf{0.840} & \textbf{0.936} \\ \hline
\end{tabular} \label{tab:ablationFusion}
\end{table*}

Using the same VGG16 backbone network, we compare different ROI in Table~\ref{tab:ablationFusion}. In the first 5 rows, single ROIs are used as input to the regression network to predict the BMDs. The first 4 rows are results using local ROIs, and the \textit{5th} row, \textit{chest}, uses the whole chest image (global ROI model). The \textit{cervical} models results in the best performance among the the local ROIs, reporting an \textit{R-value} of \textit{0.795} and an \textit{AUC} of \textit{0.906}. The \textit{chest} model performs better than all local ROI models since the whole CXR image contains the overall bone information.

\subsection{Comparison with Individual ROIs}
\label{ssec:fusionmodel}

As there are many bones in the CXR scan, the BMD of one bone may not be consistent with another. ROI models lack the mechanism to solve conflicts of inconsistent BMD from different bones. The key idea of \textit{multi-ROI model} is to allow the network to automatically identify the informative regions for estimating lumbar spine BMD. 

The multi-ROI model in Figure~\ref{fig:system} takes \textit{7} input modalities all at once. They share the same encoding backbone. Each encoded feature vector is of size \textit{512}, the concatenated feature has a shape of \textit{512*7}. Individual ROI features share the same fully connected layer, while the concatenated vector goes through a stand-along FC layer. 

We plot the predicted BMDs and ground truth BMDs in Figure~\ref{fig:fusionFClayerPredPoints}. Compared to prediction distribution of other ROI models, the feature concatenation and its following FC layer in the multi-ROI model tightens the prediction range, reducing predicting error variation. The shared FC layer enforces regularization on individual feature vector, reducing the absolute predicting error.

The multi-ROI model achieves an average R-value of \textit{0.840} and an average AUC score of \textit{0.936}, surpassing all individual ROI schemes in Table~\ref{tab:ablationFusion}. Compared with the chest model, the multi-ROI model encodes additional 6 local ROI modalities, which allows for detailed local texture computation. The feature concatenation retains encoded information from both local ROIs and the global chest, involving regional information interactions. The multi-ROI model even yields a higher AUC score than model ensembles in Table~\ref{tab:ablationEnsemble}. 

\begin{table*}[]
	\caption{Ensemble\_roi calculate the average of 5 ROI model predictions as result, Ensemble\_all use all individual model predictions as result.}
	\centering
	\begin{tabular}{|c|c|c|c|c|c|c|c|c|c|c|}
\hline
\multirow{2}{*}{Ensemble} & \multicolumn{2}{c|}{L1}         & \multicolumn{2}{c|}{L2}         & \multicolumn{2}{c|}{L3}         & \multicolumn{2}{c|}{L4}         & \multicolumn{2}{c|}{Average}    \\ \cline{2-11} 
                          & R-value        & AUC            & R-value        & AUC            & R-value        & AUC            & R-value        & AUC            & R-value        & AUC            \\ \hline\hline
Ensm\_roi             & 0.879          & 0.935          & 0.876          & 0.946          & 0.850          & 0.929          & 0.812          & 0.894          & 0.854          & 0.926          \\ \hline
Ensm\_all             & \textbf{0.881} & \textbf{0.938} & \textbf{0.880} & \textbf{0.949} & \textbf{0.851} & \textbf{0.933} & \textbf{0.816} & \textbf{0.903} & \textbf{0.857} & \textbf{0.931} \\ \hline
\end{tabular} \label{tab:ablationEnsemble}
\end{table*}

\subsection{Evaluating Model Ensemble Performance}
\label{ssec:ensemblemodel}
The individual model performance is influenced by the data distribution, input modality choice, and the training process. Overfitting is also a challenge due to the limited amount of labeled data. A common strategy to boost model performance is ensemble learning. In Table~\ref{tab:ablationFusion}, we have 6 models using different ROIs. The mean of the BMD predictions from the these models is used as ensemble predictions, listed in Table~\ref{tab:ablationEnsemble}. \textit{Ensemble\_roi} is the ensemble of all single-ROI models (\textit{cervical, clavicle, lumbar, ribcage, chest} models), and \textit{Ensemble\_all} is the ensemble of all models including the multi-ROI model.

Compared to \textit{Ensemble\_roi}, \textit{Ensemble\_all} has the additional \textit{multi-ROI model} and it performs stronger in all the metrics. Although \textit{Ensemble\_roi} has information from local modalities and the whole chest X-ray, the simple average of single-ROI model predictions lack the adaptive feature fusion process. In the \textit{multi-ROI model}, the concatenated feature and the following FC layer exploit all regional information. The different importance of various regions is learned in the FC layer weights. 

Even though both ensemble schemes work better in terms of averaged \textit{R-value}, they do not exceed the \textit{multi-ROI model} in terms of averaged \textit{AUC score}. A higher R-value indicates high consistency of sequential order with ground truth, while a higher AUC score means better osteoporosis decision around the T-score threshold. \textit{multi-ROI model} has a substantially higher AUC score than any individual ROI model, proving its reliability to judge osteoporosis status. The \textit{Ensemble\_all} could not improve \textit{multi-ROI model} upon the AUC score, meaning that averaged prediction from ROI models are less accurate around the T-score threshold.

\subsection{Evaluation of CXR BMD Prediction Task}
\label{ssec:ensemblemodel}
It is observed that adjacent bones have more consistent BMD scores, while the correlation decrease as bone distances increase. For example, in the lumbar BMD ground truth, L1 has a \textit{R-value} of \textit{0.907} to L2, and has a \textit{R-value} of \textit{0.883} to L3, but only has a \textit{R-value} of \textit{0.825} to L4. So the model performance may not exceed the neighboring lumbar BMD \textit{R-values} (such as L1 and L3 ). 

Another cause of the performance bottleneck is that there exist physiological differences between lumbar bones and chest bones. The accuracy for cross-region BMD prediction would have an upper limit due to osteoporosis distribution and bone differences. This can be observed in Table\ref{tab:ablationFusion}, where the \textit{cervical} model perform best among local ROI models, and \textit{ribcage} model performs worst. The cervical vertebra has a more similar role to the lumbar in spine activity, while bones in the rib cage ROIs stay static from body activities. The \textit{lumbar} model's low performance is due to X-ray overexposure and noise in the lumbar region of frontal-view CXR scan.

By analyzing the lumbar BMD ground truth, we find that L1 BMD is generally smaller than L2's, L2 BMD is generally smaller than L3's and L3 BMD is generally smaller than L4's. In Figure~\ref{fig:fusionFClayerPredPoints}, the L1 BMD distribution has a much lower range than L4's.
The \textit{multi-ROI} model get a \textit{R-value} of \textit{0.875} and an \textit{AUC} score of \textit{0.936} on L1 BMD prediction, which is already close to the \textit{R-value} between the ground truth BMDs of L1 and L3 (\textit{0.883}). This further proves the promising applicability of our proposed algorithm.

\section{Conclusion}
In this paper, we studied the task of BMD prediction from chest X-ray images. We introduced a multi-ROI model to jointly analyze multiple local and global ROIs from the CXR to predict the patient's lumbar spine BMDs. Experimental results and analysis show that by incorporating multiple ROIs, our model is able to produce significant improved performance compared to using individual ROIs or the whole CXR. Our method shows great potential to be applied to opportunistic screening of osteoporosis.

%
%
%
\bibliographystyle{splncs04}
\bibliography{mybibliography}

\begin{thebibliography}{1}
\providecommand{\url}[1]{\texttt{#1}}
\providecommand{\urlprefix}{URL }
\providecommand{\doi}[1]{https://doi.org/#1}

\bibitem{cheng2020opportunistic}
Cheng, X., Zhao, K., Zha, X., Du, X., Li, Y., Chen, S., Wu, Y., Li, S., Lu, Y.,
  Zhang, Y., et~al.: {Opportunistic Screening Using Low-Dose CT and the
  Prevalence of Osteoporosis in China: A Nationwide, Multicenter Study}.
  Journal of Bone and Mineral Research  (2020)

\bibitem{dagan2020automated}
Dagan, N., Elnekave, E., Barda, N., Bregman-Amitai, O., Bar, A., Orlovsky, M.,
  Bachmat, E., Balicer, R.D.: {Automated opportunistic osteoporotic fracture
  risk assessment using computed tomography scans to aid in FRAX
  underutilization}. Nature Medicine  \textbf{26}(1),  77--82 (2020)

\bibitem{jang2019opportunistic}
Jang, S., Graffy, P.M., Ziemlewicz, T.J., Lee, S.J., Summers, R.M., Pickhardt,
  P.J.: {Opportunistic osteoporosis screening at routine abdominal and thoracic
  CT: normative L1 trabecular attenuation values in more than 20 000 adults}.
  Radiology  \textbf{291}(2),  360--367 (2019)

\bibitem{lewiecki2019challenges}
Lewiecki, E.M., Leader, D., Weiss, R., Williams, S.A.: {Challenges in
  osteoporosis awareness and management: results from a survey of US
  postmenopausal women}. Journal of Drug Assessment  \textbf{8}(1),  25--31
  (2019)

\bibitem{li2020structured}
Li, W., Lu, Y., Zheng, K., Liao, H., Lin, C., Luo, J., Cheng, C.T., Xiao, J.,
  Lu, L., Kuo, C.F., et~al.: {Structured landmark detection via
  topology-adapting deep graph learning}. arXiv preprint arXiv:2004.08190
  (2020)

\bibitem{pickhardt2020automated}
Pickhardt, P.J., Graffy, P.M., Zea, R., Lee, S.J., Liu, J., Sandfort, V.,
  Summers, R.M.: {Automated abdominal CT imaging biomarkers for opportunistic
  prediction of future major osteoporotic fractures in asymptomatic adults}.
  Radiology  \textbf{297}(1),  64--72 (2020)

\bibitem{smith2019screening}
Smith, A.D.: {Screening of bone density at CT: an overlooked opportunity}
  (2019)

\bibitem{sozen2017overview}
S{\"o}zen, T., {\"O}z{\i}{\c{s}}{\i}k, L., Ba{\c{s}}aran, N.{\c{C}}.: {An
  overview and management of osteoporosis}. European journal of rheumatology
  \textbf{4}(1), ~46 (2017)

\end{thebibliography}
%

\end{document}


%


\begin{figure*}[bt!]
	\centering
	\includegraphics[width=0.95\linewidth]{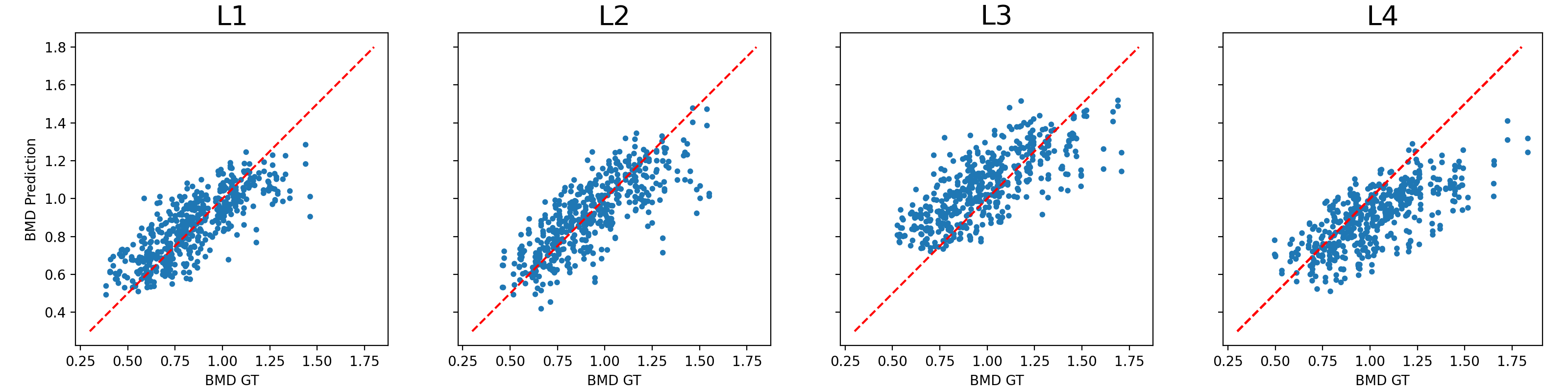}
	\caption{Clavicle model BMD prediction on different lumbar vertebra (L1-L4). Local ROI model prediction is more loosely distributed.}
	\label{fig:fusionFClayerPredPoints}
\end{figure*}

\begin{figure*}[bt!]
	\centering
	\includegraphics[width=0.95\linewidth]{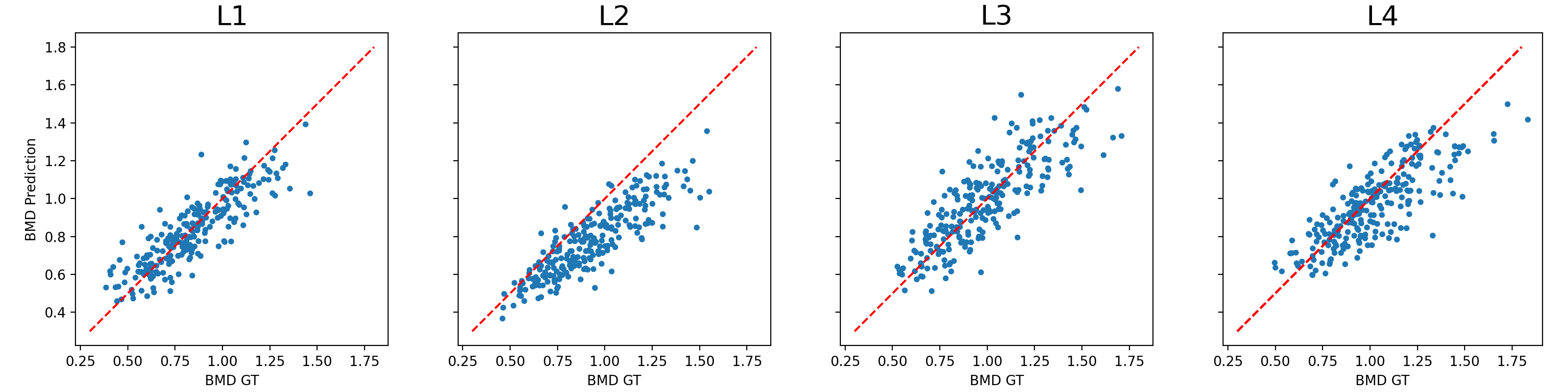}
	\caption{Chest model. The whole chest contain global information, and the chest model prediction is more robust to noise.}
	\label{fig:fusionFClayerPredPoints}
\end{figure*}

\begin{figure*}[bt!]
	\centering
	\includegraphics[width=0.95\linewidth]{Figures/Comb_testDataBMD_fusion wo. ROI FC layer.png}
	\caption{Multi-ROI model without the shared ROI FC layer. The prediction range is reduced by the concatenated feature and its following FC layer.}
	\label{fig:fusionFClayerPredPoints}
\end{figure*}

\begin{figure*}[bt!]
	\centering
	\includegraphics[width=0.95\linewidth]{Figures/Comb_testDataBMD_fusion w. ROI FC layer.png}
	\caption{Multi-ROI model with the shared ROI FC layer. The FC layer regulates individual ROI feature vectors, and reduces BMD error.}
	\label{fig:fusionFClayerPredPoints}
\end{figure*}